\documentstyle{aipproc}

\newcommand{\beq}{\begin{equation}}
\newcommand{\eeq}{\end{equation}}
\newcommand{\beqa}{\begin{eqnarray}}
\newcommand{\eeqa}{\end{eqnarray}}
\newcommand{\noi}{\noindent}

\newcommand{\lsim}{\mathrel{\lower4pt\hbox{$\sim$}}
\hskip-12.5pt\raise1.6pt\hbox{$<$}\;}

\newcommand{\gsim}{\mathrel{\lower4pt\hbox{$\sim$}}
\hskip-12.5pt\raise1.6pt\hbox{$>$}\;}

\newcommand{\amzms}{\alpha(m_Z)_{\overline{MS}}}
\newcommand{\aomzms}{\alpha^{-1}(m_Z)_{\overline{MS}}}
\newcommand{\mzms}{(m_Z)_{\overline{MS}}}
\newcommand{\ssthw}{\sin^2\theta_W}
\newcommand{\ssthwz}{\sin^2\theta^0_W}
\newcommand{\ssthweff}{\sin^2\theta^{\rm eff}_W}

\begin{document}

\begin{flushright}
BNL-HET-00/04 \\
hep-ph/000 \\ March 2000
\end{flushright}

\title{Precision Electroweak Parameters  and the Higgs Mass\thanks{To
be published in the Proceedings of MuMu99-5th 
International Conference on Physics Potential and Development of
$\mu^\pm\mu^-$ Colliders, San Francisco, CA, Dec.\ (1999)}}

\author{William J. Marciano} 
\address{Brookhaven National Laboratory\thanks{This manuscript has been authored
under contract number DE-AC02-98CH10886 with the U.S. Department of
Energy. Accordingly, the U.S. Government retains a non-exclusive,
royalty-free license to publish or reproduce the published form of this
contribution, or allow others to do so, for U.S. Government
purposes.} \\
Upton, New York\ \ 11973 \\}

\maketitle
\begin{abstract}%
 The status of various precisely measured electroweak parameters is
 reviewed. Natural relations among them are shown to constrain the
 Higgs mass, $m_H$, via quantum loop effects to relatively low values.
 A comparison with direct Higgs searches is made.
\end{abstract}

\section*{Fundamental Parameters and Precision Measurements}

The SU(2)$_L\times{}$U(1)$_Y$ electroweak sector of the standard model
contains 17 or more fundamental parameters. They include gauge and
Higgs field couplings as well as fermion masses and mixing angles. In
terms of those parameters, predictions can be made with high accuracy
for essentially any electroweak observable. Very precise measurements
of those quantities can then be used to test the standard model, 
at the quantum loop level and predict the Higgs scalar mass or search for
small deviations from expectations which would indicate ``New Physics''.

Some fundamental electroweak parameters have been determined with
extraordinary precision. Foremost in that category is the fine
structure constant $\alpha$. It is  best obtained by comparing the
measured \cite{prl5926} anomalous magnetic moment of the electron,
$a_e\equiv (g_e-2)/2$

\beq
a^{\rm exp}_e = 1159652188 (3) \times 10^{-12} \label{eq1}
\eeq

\noi with the calculated 4 loop QED prediction \cite{hepph9810512}

\beqa
a^{\rm th}_e & = & \frac{\alpha}{2\pi} - 0.328478444
\left(\frac{\alpha}{\pi}\right)^2 + 1.181234 \left(\frac{\alpha}{\pi}
\right)^3 - 1.5098 \left(\frac{\alpha}{\pi}\right)^4 \nonumber \\
& & \quad +1.66\times 10^{-12} \label{eq2}
\eeqa

\noi where the $1.66\times10^{-12}$ comes from small hadronic and weak
loop effects. Assuming no significant ``new physics'' contributions to
$a^{\rm th}_e$, it can be equated with (\ref{eq1}) to give

\beq
\alpha^{-1} = 137.03599959 (40) \label{eq3}
\eeq

\noi That precision is already very impressive.  Improvement
by a factor of 10 appears to be technically feasible \cite{gabrielse} and should
certainly be undertaken. However, at this time such improvement would
not further our ability to test QED\null. Pure  QED tests require comparable
measurements of $\alpha$ in other processes. Agreement between two
distinct $\alpha$ determinations tests QED and probes for ``new
physics'' effects. After $a_e$, the next best (direct) measurement of
$\alpha$ comes from the quantum Hall effect

\beq
\alpha^{-1} (qH) = 137.03600370 (270) \label{eq4}
\eeq

\noi which is not nearly as precise. Nevertheless, the agreement of
(\ref{eq3}) and (\ref{eq4}) (at the 1.50 sigma level) is a major triumph for
QED up to the 4 loop quantum level.

In terms of probing ``new physics'', one can search for a shift in
$a_e$ by $m^2_e/\Lambda^2_e$ where $\Lambda_e$ is the approximate scale
of some generic new short-distance effect. Current comparison of
$a_e\to \alpha$ and $\alpha(qH)$ explores $\Lambda_e\lsim 100$
GeV\null. To probe the much more interesting $\Lambda_e\sim{\cal O}$
(TeV) region would require an order of magnitude improvement in $a_e$
and about two orders of magnitude error reduction in some direct
precision determination of $\alpha$ 
such as the quantum Hall effect. Perhaps the most likely possibility is
to use the already very precisely measured Rydberg constant in
conjunction with a much improved $m_e$ determination to obtain an
independent $\alpha$.

The usual fine structure constant, $\alpha$, is defined at zero
momentum transfer as is appropriate for low energy atomic physics
phenomena. However, that definition is not well suited for
short-distance electroweak effects. Vacuum polarization loops
screen charges such that the effective (running) electric charge increases at
short-distances. One can incorporate those quantum loop contributions
into a short-distance \cite{prd20274} $\alpha(m_Z)$ defined at
$q^2=m^2_Z$. The main effect comes from lepton loops, which can be very
precisely calculated, and somewhat smaller hadronic loops. The latter
are not as
theoretically clean and must be obtained by combining perturbative
calculations with results of a dispersion relation which employs
${\cal O}(e^+e^-\to{}$hadrons) data. A detailed study by Davier and
H\"ocker found \cite{plb439427}

\beq
\alpha^{-1}(m_Z) = 128.933 (21) \label{eq5}
\eeq

\noi where the uncertainty stems primarily from low energy hadronic loops. Although not
nearly as precise as $\alpha^{-1}$, the uncertainty quoted in
(\ref{eq5}) is impressively small and a tribute to the effort that has
gone into reducing it. (When I first studied this issue in 1979, I
crudely estimated \cite{prd20274} $\alpha^{-1}(m_Z)\simeq 128.5\pm1.0$.)
However, the error in (\ref{eq5}) is still somewhat controversial,
primarily because of its reliance on perturbative QCD down to very low
energies. For comparison, an earlier study by Eidelman and
Jegerlehner \cite{zpc67585}, which relied less on perturbative QCD and
more on $e^+e^-$ data found

\beq
\alpha^{-1}(m_Z) = 128.896 (90) \qquad ({\rm E~\&~J~1995}) \label{eq6}
\eeq

\noi That estimated uncertainty is often cited as more conservative and
therefore sometimes employed in $m_H$ and ``new physics'' constraints.
As we shall see, the 
smaller uncertainty in (\ref{eq5}) has very important consequences for
predicting the Higgs mass. I note that a more recent study \cite{jeger}
by Eidelman and Jegerlehner finds

\beq
\alpha^{-1}(m_Z) = 128.913 (35) \qquad ({\rm E~\&~J~1998}) \label{eq7}
\eeq

\noi which is in good accord with (\ref{eq5}) and also exhibits
relatively small uncertainty. In my subsequent
discussion, I employ the result in (\ref{eq5}), but caution the reader
that a more conservative approach would expand the uncertainty, perhaps
even by as much as a factor of 4.

A related short-distance coupling, $\amzms$, can be defined by modified
minimal subtraction at scale $\mu=m_Z$. It is particularly useful for
studies of coupling unification in grand unified theories (GUTS) where
a uniform comparitive definition $(\overline{MS})$ of all couplings is called
for \cite{prl46163}. The quantities $\alpha(m_Z)$ and $\amzms$ differ by
a constant, such that \cite{hepph9803453}

\beq
\aomzms = \alpha^{-1}(m_Z) - 0.982 = 127.951 (21) \label{eq8}
\eeq

In weak interaction physics, the most precisely determined parameter is
the Fermi constant, $G_\mu$, as obtained from the muon lifetime. One
extracts that quantity by comparing the experimental value 

\beq
\tau_\mu = 2.197035 (40) \times10^{-6} s \label{eq9}
\eeq

\noi with the theoretical prediction

\beqa
\tau^{-1}_\mu = \Gamma(\mu\to{\rm all}) & = & \frac{G^2_\mu
m^5_\mu}{192\pi^3} f \left(\frac{m^2_e}{m^2_\mu}\right) (1+{\rm R.C.})
\left(1+ \frac{3}{5}\, \frac{m^2_\mu}{m^2_W}\right) \nonumber \\
f(x) & = & 1-8x+8x^3 -x^4 -12x^2\ell nx \label{eq10} 
\eeqa

\noi In that expression R.C. stands for Radiative Corrections. Those
terms are somewhat arbitrary in the standard model. The point being that
$G_\mu$ is a renormalized parameter which is used to absorb most loop
corrections to muon decay. Those corrections not absorbed into $G_\mu$
are explicitly factored out in R.C\null. For historical reasons and in
the spirit of effective field theory approaches, R.C. has been chosen
to be the QED corrections to the old V-A four fermion description of  muon
decay \cite{ap2020}. That definition is practical, since the QED
corrections to muon decay in the old V-A theory are finite to all
orders in perturbation theory. In that way, one finds

{\footnotesize
\beq
{\rm R.C.} = \frac{\alpha}{2\pi} \left(\frac{25}{4}-\pi^2\right) \left(
1+\frac{\alpha}{\pi} \left(\frac{2}{3}\ell n \frac{m_\mu}{m_e} -3.7
\right) + \left(\frac{\alpha}{\pi}\right)^2 \left(\frac{4}{9} \ell n^2
\frac{m_\mu}{m_e} - 2.0 \ell n \frac{m_\mu}{m_e} +C\right)\cdots\right)
\label{eq11} 
\eeq}

\noi The leading ${\cal O}(\alpha)$ terms in that expression have been known for
a long time from the pioneering work of Kinoshita and
Sirlin \cite{pr1131652} and Berman \cite{pr112267}. Coefficients of the
higher order logs can be obtained from the renormalization group
constraint \cite{npb29296} 

\beqa
&& \left(m_e \frac{\partial}{\partial m_e} + \beta(\alpha)
\frac{\partial}{\partial\alpha} \right) {\rm R.C.} =0 \nonumber \\
&& \quad \beta(\alpha) = \frac{2}{3}\,\frac{\alpha^2}{\pi} +
\frac{1}{2}\, \frac{\alpha^3}{\pi^2} \cdots \label{eq12}
\eeqa

\noi The -3.7 two loop constant in parenthesis was recently computed
by van Ritbergen and Stuart \cite{prl82488}. It almost exactly cancels
the leading log two loop correction obtained from the renormalization group
approach (or mass singularities argument) of Roos and
Sirlin \cite{npb29296}. Hence, the original ${\cal O}(\alpha)$ correction in
(\ref{eq9}) is a much better approximation than one might have
guessed. Comparing (\ref{eq9}) and (\ref{eq10}), one finds

\beq
G_\mu = 1.16637 (1)\times10^{-5}{\rm~GeV}^{-2} \label{eq13}
\eeq

There have been several experimental proposals \cite{czar} to reduce the
uncertainty in $\tau_\mu$ and $G_\mu$ by as much as  a factor of 20. Such
improvement appears technically feasible and, given the fundamental
nature of $G_\mu$, should certainly be undertaken. However, from the
point of view of testing the standard model, the situation is similar
to $\alpha$. $G_\mu$ is already much better known than the other parameters
it can be compared with; so, significant improvement must be made in
other quantities before a more precise $G_\mu$ is required. 

Let me emphasize the fact that lots of interesting loop effects have
been absorbed into the renormalization of
$g^2_{2_0}/4\sqrt{2}m^{0^2}_W$ which we call $G_\mu$. Included are top
quark \cite{npb12389} and Higgs loop corrections \cite{npb84132} to the
$W$ boson propagator as well as potential ``new physics'' from SUSY
loops, Technicolor etc. Even tree level effects of possible more
massive gauge bosons such as excited $W^{\ast^\pm}$ bosons could be  effectively
incorporated into $G_\mu$. To uncover those contributions requires
comparison of $G_\mu$ with other precisely measured electroweak
parameters which have different quantum loop (or tree level)
dependences. Of course, those quantities must be related to $G_\mu$ in
such a way that short-distance divergences cancel in the comparison.

Fortunately, due to an underlying global SU(2)$_V$ symmetry in the
standard model, there exist natural relations among various bare
parameters \cite{nc16a423} 

\beq
\ssthwz = \frac{e^2_0}{g^2_{2_0}} = 1- (m^0_W/m^0_Z)^2
\label{eq14}
\eeq

\noi Each of those bare unrenormalized  expressions contains
short-distance infinities; however, because the theory is
renormalizable, the divergences  are the same. 
Therefore, those 
relations continue to hold for renormalized quantities, up to finite,
calculable radiative corrections \cite{nc16a423}. The residual
radiative corrections contain very interesting effects such as $m_t$
and $m_H$ dependence as well as possible ``new physics''. So, for
example, one can relate

\beq
G_\mu = \frac{\pi\alpha}{\sqrt{2} m^2_W (1-m^2_W/m^2_Z)} (1+
rad.~corr.) \label{eq15}
\eeq

\noi and test the predicted radiative corrections, if $m_Z$ and $m_W$
are also precisely known.

Gauge boson masses are not as well determined as $G_\mu$, but they have
reached  high levels of precision. In particular, the $Z$ mass has
been measured with high statistics Breit-Wigner fits to the $Z$
resonance at LEP with the result \cite{swartz}

\beq
m_Z = 91.1871 (21)~{\rm GeV} \label{eq16}
\eeq

\noi That determination is so good that one must be very precise regarding
the definition of $m_Z$. (Remember the $Z$ has a relatively large width
$\sim2.5$ GeV\null.) The quantity in (\ref{eq16}) is related to the
real part of the $Z$ propagator pole, $m_Z$ (pole), and full width,
$\Gamma_Z$, by \cite{prl672127}

\beq 
m^2_Z = m^2_Z({\rm pole}) + \Gamma^2_Z \label{eq17}
\eeq

\noi The two mass definitions $m_Z$ and $m_Z$ (pole) differ by about 34
MeV, which is much larger than the uncertainty in (\ref{eq16}). Hence,
one must specify which definition is being employed in precision
studies. I note, that the $m_Z$ in (\ref{eq16}) is also more appropriate for
use in low energy neutral current amplitudes.

In the case of the $W^\pm$ bosons, the renormalized mass, $m_W$, is
similarly defined by 

\beq
m^2_W = m^2_W ({\rm pole}) + \Gamma^2_W \label{eq18}
\eeq

\noi That quantity is obtained \cite{lanc} from studies at $p\bar p$ colliders,
$m_W=80.448 (62)$ GeV, as well as $e^+e^-\to W^+W^-$ at LEPII, $m_W=80.401
(48)$ GeV\null. Together they average to

\beq
m_W=80.419 (38)~{\rm GeV}\;. \label{eq19}
\eeq

\noi The current level of uncertainty, $\pm 38$ MeV, is large compared
to $\Delta m_Z$. It is expected that continuing efforts at LEPII and
Run II at Fermilab's Tevatron should reduce that error to about $\pm25$
MeV\null. A challenging but worthwhile goal for future high energy
facilities would be to push $\Delta m_W$ to $\pm10$ MeV or better. At
that level, all sorts of interesting ``new physics'' effects are
probed. I  note that the $m_W$ defined in (\ref{eq19}) 
is also the appropriate quantity for low energy amplitudes such as muon
decay.

Another important quantity for precision standard model tests is $m_t$, the
top quark mass. Measurements from CDF and D$\emptyset$ at Fermilab give \cite{lanc}

\beq
m_t ({\rm pole}) = 174.3\pm5.1~{\rm GeV} \label{eq20}
\eeq

\noi Reducing that uncertainty further is important for quantum loop
studies, as we shall subsequently
see. Future Tevatron Run II efforts are expected to reduce the error in
$m_t$ to about $\pm2$ GeV\null. LHC and NLC studies should bring it
well below $\pm1$ GeV\null.

In addition to masses, the renormalized weak mixing angle plays a
central role in tests of the standard model. That parameter can be
defined in a variety of ways, each of which has its own advocates. I
list three popular examples \cite{prd20274,pr491160,czartwo}

\beqa
&\ssthw\mzms & \qquad (\overline{MS} {\rm ~definition~at~}\mu=m_Z)
\qquad\qquad\qquad\,\,\,\,\, (a) \nonumber \\
&\ssthweff &  \qquad (Z\mu\bar\mu{\rm ~vertex})
\qquad\qquad\qquad\qquad\qquad\qquad\quad\, (b) \label{eq21} \\
& \ssthw\equiv &\!\!\!\!\!\!\! 1-m^2_W/m^2_Z
\qquad\qquad\qquad \qquad\quad\qquad\qquad\qquad\qquad\!\!
(c) \nonumber
\eeqa

\noi They differ by finite ${\cal O}(\alpha)$ loop corrections. The
$\overline{MS}$ definition is particularly simple, being defined as the
ratio of two $\overline{MS}$ couplings $\ssthw\mzms\equiv
e^2\mzms /$ $g^2_2 \mzms$. It was introduced for GUT
studies \cite{prl46163}, but is useful for most electroweak analyses.
The effective, $\ssthweff$, weak angle was invented for
$Z$ pole analyses. Roughly speaking, it is defined by the ratio of
vector and axial-vector components (including loops) for the
on-mass-shell $Z\mu\bar\mu$ vertex${}\to1$--$4\ssthweff$. Although
conceptually rather  simple, analytic electroweak 
radiative corrections expressed in terms of $\ssthweff$
are complicated and ugly. Numerically, it is close to the
$\overline{MS}$ definition \cite{pr491160}

\beq
\ssthweff = \ssthw\mzms + 0.00028 \label{eq22}
\eeq

\noi but the analytic structure of the difference is quite complicated. For
those intent on employing $\ssthweff$, a strategy might
be to calculate radiative corrections in terms of $\ssthw$ $\mzms$
and then translate to $\ssthweff$ via (\ref{eq22}). But
why not simply use $\ssthw$ $\mzms$?

Currently, $Z$ pole studies at LEP and SLAC give \cite{swartz}

\beqa 
\ssthw\mzms & = & 0.23091\pm 0.00021 \nonumber \\
\ssthweff & = & 0.23119 \pm 0.00021 \label{eq23}
\eeqa

\noi That result includes  measurements of the
left-right asymmetry, $A_{LR}$, at SLAC as well as the various lepton
asymmetries at LEP and SLAC\null. The $A_{LR}$ contribution has for
some time given a relatively low value for the weak mixing angle. The
latest \cite{swartz} SLD result is

\beq
\sin^2\theta_W(m_Z)_{\overline{MS}} = 0.23073 (28) \label{eq24}
\eeq

\noi Currently, the $Z\to b\bar b$ forward-backward asymmetry at LEP gives a
higher $\ssthweff$ and, if included,  would bring up the average.
However, the $Zb\bar b$ coupling appears to be
somewhat anomalous and suggests problematic $b$ identification; so, one should be cautious when including such
results in averages. 

There are very good reasons to  clarify and further improve
$\sin^2\theta_W(m_Z)_{\overline{MS}}$. One could imagine redoing
$A_{LR}$ at a future polarized lepton-lepton ($e^+e^-$ or $\mu^+\mu^-$) collider,
but with much higher statistics. In principle, one might reduce
the uncertainty in $\ssthweff$ by a factor of 10 to $\pm0.00002$, an incredible achievement if
accomplished \cite{czartwo}.

The so-called on-shell or mass definition \cite{prd22971} in
(\ref{eq21}c) also has its advocates. It can be directly obtained from $m_W$ and
$m_Z$ determinations. Indeed, at hadron colliders, the ratio $m_W/m_Z$
can have reduced systematic uncertainties. One could imagine that the
current uncertainty in 

\beq
\ssthw = 1-m^2_W/m^2_Z = 0.2222\pm0.0007 \label{eq25}
\eeq

\noi might be reduced by a factor of about 4 at the LHC\null. Such a
reduction is extremely important since the comparison of $\ssthw$ and
$\ssthw\mzms$ provides a clean probe of ``new physics''. It is also
possible (because of a subtle cancellation of certain loop
effects \cite{npb189442}) to measure $\ssthw$ more directly in deep-inelastic $\nu_\mu
N$ scattering. Indeed, a recent Fermilab experiment
found \cite{epjc1509}

\beq
\ssthw = 0.2255\pm0.0019\pm0.0010 \label{eq26}
\eeq

\noi where the first error is statistical and the second systematic.
That single measurement is quite competitive with (\ref{eq25}) and
complements it nicely. One might imagine a future high statistics
effort significantly reducing the error in (\ref{eq26}), but that would
require a new, intense,  high energy neutrino beam.

Two other well measured electroweak parameters are the charged and
neutral leptonic partial widths of the $Z$ boson \cite{swartz}

\beqa
\Gamma (Z\to\ell^+\ell^-(\gamma)) & = & 83.96\pm0.09 {\rm ~MeV}
\nonumber \\
\Gamma (Z\to\Sigma\nu\bar\nu) & = & 498.8\pm1.5 {\rm ~MeV} \label{eq27}
\eeqa

\noi The first of those, by definition, corresponds to $Z$ decay into
massless charged leptons along with the possibility of inclusive
bremsstrahlung. The second represents the inclusive invisible width of
the $Z$.

All of the above precision measurements can be collectively used to
test the standard model, predict the Higgs mass, and search for ``new
physics'' effects. That ability stems from the natural relations in
(\ref{eq14}) and calculations \cite{prd22971,prd222695} of the radiative
corrections to them. Parametrizing those radiative corrections by
$\Delta r$, $\Delta r\mzms$, and $\Delta\hat r$, one
finds \cite{npb35149}

\beqa
\frac{\pi\alpha}{\sqrt{2}G_\mu m^2_W} & = &
\left(1-\frac{m^2_W}{m^2_Z}\right) (1-\Delta r)
\qquad\qquad\qquad\qquad \qquad\qquad\! (a) \nonumber \\
\frac{\pi\alpha}{\sqrt{2}G_\mu m^2_W} & = & \ssthw\mzms (1-\Delta
r\mzms) \qquad\qquad\qquad\quad (b) \label{eq28} \\
\frac{4\pi\alpha}{\sqrt{2}G_\mu m^2_Z} & = & \sin^22\theta_W\mzms
(1-\Delta\hat r) \qquad\qquad\qquad\qquad\qquad (c) \nonumber
\eeqa

\noi Those expressions contain all one loop corrections to $\alpha$,
muon decay, $m_W$, $m_Z$ and $\ssthw\mzms$ and incorporate some leading two loop
contributions. The quantities $\Delta r$ and $\Delta\hat r$ are
particularly interesting because of their dependence on $m_t$ and
$m_H$. In addition, all three quantities provide probes of ``new
physics''.

Numerically, all three radiative corrections in (\ref{eq28}) contain a
significant contribution from vacuum polarization
effects \cite{prd20274} in $\alpha$, 
about $+7\%$. They are basically the same as the corrections that enter
into the evolution of $\alpha$ to $\alpha(m_Z)$. Leptonic loops
contribute a significant part of that effect and can be very accurately
computed. Hadronic loops are less clean theoretically and lead to a
common uncertainty in $\Delta r$, $\Delta r \mzms$, and $\Delta\hat r$
of

\beq
-\alpha \Delta\alpha^{-1}(m_Z) \label{eq29}
\eeq

\noi For $\Delta\alpha^{-1}(m_Z) = 0.021$ as in (\ref{eq5}), that
amounts to a rather negligible $\pm0.00015$ error. However, for
$\Delta\alpha^{-1}(m_Z)=\pm0.090$ as in (\ref{eq6}), it increases to
$\pm0.00066$. That large an uncertainty would impact precision tests.
If one wishes to avoid that low energy hadronic loop uncertainty,
dependence on $\alpha$ can be circumvented by considering

\beq
\ssthw\mzms = \left(1-\frac{m^2_W}{m^2_Z}\right) (1-\Delta r + \Delta
r\mzms) \label{eq30}
\eeq

\noi Currently, that comparison suggests a very light Higgs, but is not
yet quite competitive with eq.~(\ref{eq28}c) in constraining
$m_H$. However, future  improvements in $m_W$ and $m_t$ could make it
very interesting.

Another useful relation that provides some sensitivity to $m_H$
without $\Delta\alpha$ uncertainties involves the partial $Z$ width
when written as \cite{marc}

\beq
\Gamma(Z\to\ell^+\ell^-(\gamma)) = \frac{G_\mu m^3_Z (1-4\sin^2\theta_W
(m_Z)_{\overline{MS}}+8 \sin^4\theta_W(m_Z)_{\overline{MS}})}{12
\sqrt{2} \pi (1-\Delta r_Z(m_H))} \label{eq31}
\eeq

Using $m_t=174.3\pm5.1$ GeV as input, one can compute the radiative
corrections in (\ref{eq28}) as functions of $m_H$. Those results are
illustrated in table~\ref{tabone}. Note that $\Delta r$ is most
sensitive to changes in $m_H$ but also carries the largest uncertainty
from $\Delta m_t=\pm5.1$ GeV ($\pm0.0020$). Hence, efforts to determine $m_H$
from $m_W$ are starting to require a better measurement of $m_t$. On the other
hand, determining $m_H$ from $\ssthw\mzms$ via $\Delta\hat r$ is less
sensitive to $\Delta m_t$ but more sensitive to $\Delta
\alpha^{-1}(m_Z)$. Those dependences as well as
$\Gamma(Z\to\ell^+\ell^-(\gamma))$  are illustrated by the following
approximate relations \cite{plb418188} which are very accurate up to
$m_H\sim {\cal O}$(400 GeV)

{\small
\beqa
m_W\! & = &\! (80.385\pm0.032 \pm0.003{\rm ~GeV})\left(1- 0.00072\ell n
\left(\frac{m_H}{100{\rm ~GeV}}\right)\right. \nonumber \\
& & \qquad \left.- 1\times10^{-4} \ell n^2
\left(\frac{m_H}{100{\rm ~GeV}}\right) \right) \label{eq32} \\
\ssthw\mzms \!& = & \!(0.23112\pm0.00016\pm0.00006) \left(1+0.00226 \ell
n\left(\frac{m_H}{100{\rm ~GeV}} \right)\right) \nonumber \\
\Gamma(Z\to\ell^+\ell^-(\gamma)) & = & (84.011\pm0.047\pm0.0028{\rm ~MeV})
(1-0.00064\ell n\left(\frac{m_H}{100{\rm ~GeV}}\right)\nonumber \\
& & \qquad -0.00026 \ell
n^2 \left(\frac{m_H}{100{\rm ~GeV}}\right) \nonumber
\eeqa}

\noi where the errors correspond to $\Delta m_t=\pm5.1$ GeV and
$\Delta\alpha^{-1}(m_Z) = \pm0.021$ respectively. Note that increasing
$\Delta\alpha^{-1}(m_Z)$ to $\pm0.090$ would significantly compromise
the utility of $\ssthw\mzms$ for determining $m_H$ but have less of an
impact on $m_W$. Predictions for $m_W$ and $\ssthw\mzms$ are illustrated
in table~\ref{tabtwo} for various $m_H$ values \cite{plb394188}.

\begin{table}[tbh]
\begin{center}
\caption{Values of $\Delta r$, $\Delta r\mzms$, and $\Delta\hat r$ for
various $m_H$. A top quark mass of $174.3\pm5.1$ GeV and
$\alpha^{-1}(m_Z) = 128.933 (21)$ are assumed. \label{tabone}}
\begin{tabular}{lccc}
$m_H$ (GeV) & $\Delta r$ & $\Delta r\mzms$ & $\Delta\hat r$ \\
& $\pm0.0020\pm0.0002$ & $\pm0.0001\pm0.0002$ & $\pm0.0005\pm0.0002$ \\
\\
~75 & 0.03402 & 0.06914 & 0.05897 \\
100 & 0.03497 & 0.06937 & 0.05940 \\
125 & 0.03575 & 0.06955 & 0.05974 \\
150 & 0.03646 & 0.06964 & 0.06000 \\
200 & 0.03759 & 0.06980 & 0.06042 \\
400 & 0.04065 & 0.07005 & 0.06144 \\
\end{tabular}
\end{center}
\end{table}

\begin{table}[tbh]
\begin{center}
\caption{Predictions for $m_W$ and $\ssthw\mzms$ for various $m_H$
values. \label{tabtwo}}
\begin{tabular}{lcc}
$m_H$ (GeV) & $m_W$ (GeV) & $\ssthw\mzms$ \\ \\
~75 & 80.401 & 0.23097 \\
100 & 80.385 & 0.23112 \\
125 & 80.372 & 0.23124 \\
150 & 80.360 & 0.23133 \\
200 & 80.341 & 0.23148 \\
400 & 80.289 & 0.23184 \\
\end{tabular}
\end{center}
\end{table}

Employing $m_W =80.419 (38)$ GeV, $\ssthw\mzms=0.23091 (21)$, and
$\Gamma(Z\to \ell^+\ell^-(\gamma))=83.96 (9)$ MeV one finds from eq~(\ref{eq32})

\beqa
m_H & = & 53^{+77}_{-40}{\rm ~GeV} \qquad ({\rm from~}m_W)
\label{eq33} \\
m_H & = & 67^{+45}_{-27}{\rm ~GeV} \qquad ({\rm
from~}\ssthw\mzms) \label{eq34} \\
m_H & = & 208^{+340}_{-180} {\rm ~GeV} \qquad ({\rm
from~}\Gamma(Z\to\ell^+\ell^-(\gamma)) \label{eq35}
\eeqa

\noi Several features of those predictions are revealing. The first is that
$\ssthw$ $\mzms$ currently gives the best determination of $m_H$.
Note, however, the uncertainties scale as the central value; so, the
relatively small value, 67 GeV, helps reduce the uncertainties. Also, a
larger $\Delta\alpha^{-1}(m_Z)=\pm0.090$ would significantly increase
the overall uncertainty \cite{hepph9812332}. In the case of $m_W$, a
light Higgs is also suggested. In fact, the $m_W$ and
$\sin^2\theta_W(m_Z)_{\overline{MS}}$ determinations of $m_H$ are very
consistent. 

Taken together, (\ref{eq33}) and (\ref{eq34}) appear to point to a
relatively light Higgs scalar that cannot be too far from the current
LEPII bound based on the non observation of $e^+e^-\to ZH$ \cite{wu}

\beq
m_H>106 {\rm ~GeV} \qquad  \label{eq36}
\eeq

\noi  In the near future, searching for the Higgs via associated
$W^\pm H$ and $ZH$ at the Fermilab $p\bar p$ collider during Run II
promises discovery up to $m_H\sim 115$--130 GeV, perhaps even higher. Higgs
unveiling may soon be at hand.

\section*{Discussion}

Within the Standard Model framework, precision electroweak studies
suggest a relatively light Higgs scalar. The upper bound from global
fits to all data is about \cite{erler} 235 GeV\null. Such a bound is in
accord  with ``no new physics'' scenarios, where perturbative validity up
to ${\cal O}(10^{19}$ GeV) and vacuum stability imply the range \cite{sirlin}

\beq
135 {\rm ~GeV} \lsim m_H\lsim 180 {\rm ~GeV} \label{eq37}
\eeq

Supersymmetry, on the other hand, is starting to be squeezed by lack of
a Higgs scalar discovery at LEP II\null. The MSSM scenario requires
$m_H\lsim 135$ GeV, but one generally finds that considerably lower values
are preferred. The current bound of $m_H\gsim 106$ GeV will be pushed
to $\sim 112$ GeV at LEP II and somewhat further by Run II at the
Fermilab Tevatron. If MSSM has some relation to Nature's truth, a
discovery should not be far off.

What about alternative theories where there is no fundamental Higgs? In
some dynamical electroweak symmetry breaking scenarios, one expects an
effective 
$m_H\sim {\cal O}$(1 TeV). Are such ideas ruled out? They face a basic
difficulty which is nicely illustrated by the $S$ \& $T$ parameters of
Peskin and Takeuchi \cite{peskin}. For $m_H\simeq 1$ TeV, one finds the
following predictions 

\beqa
m_W & \simeq & 80.214 {\rm ~GeV} \quad (1-0.0036S + 0.0056T) \nonumber
\\
\sin^2\theta_W(m_Z)_{\overline{MS}} & \simeq & 0.23250 \quad (1+0.0158S
- 0.0113T) \label{eq38} \\
\Gamma(Z\to\ell^+\ell^-(\gamma)) & \simeq & 83.79 {\rm ~MeV} \quad
(1-0.0021S + 0.0093T) \nonumber
\eeqa

\noi where $S$ and $T$ represent loop effects from heavy, strongly
interacting, condensate fermions. Comparison of $m_W$ and
$\sin^2\theta_W(m_Z)_{\overline{MS}}$  via \cite{marctwo}

\beq
S\simeq 118 \left\{ 2\left( \frac{m_W-80.214{\rm ~GeV}}{80.214 {\rm
~GeV}} \right) + \frac{\sin^2\theta_W(m_Z)_{\overline{MS}} -
0.23250}{0.23250} \right\} \label{eq39}
\eeq

\noi suggests from $m_W = 80.419 (38)$ GeV and
$\sin^2\theta_W(m_Z)_{\overline{MS}} = 0.23091 (21)$

\beq
S\simeq - 0.20\pm0.15 \label{eq40}
\eeq

\noi while consistency with eq.~(\ref{eq38}) requires

\beq
T\simeq 0.33\pm0.17 \label{eq41}
\eeq

\noi Generating $T\simeq0.3$ by loop effects is actually quite easy. It
requires  relatively small mass splittings of new heavy fermion
isodoublets. The problem is a negative $S$ which is possible, but not 
particularly natural in simple dynamical models where $S\sim{\cal
O}(1)$ is more likely. However, one could easily imagine small shifts
in $m_W$ and $\sin^2\theta_W(m_Z)_{\overline{MS}}$ which could give
$S\simeq 0$ and $T>0$. So, $m_H\simeq 1$ TeV
could be accommodated by non-vanishing $S$ \& $T$, but it is much easier
to simply satisfy the precision measurement constraints with $m_H\simeq
100$--200 GeV and $S\simeq T\simeq 0$.

\end{document}